\documentstyle[aps,prl,epsfig,multicol]{revtex}

\begin{document}

\title{Measurement of the Induced Proton Polarization $\bf{P_n}$\\
in the \bf{$\rm ^{12} C$($ {\rm\bf e}$,$\rm\bf e^\prime{\rm\bf\vec p}\,$)} 
Reaction}
\author{R. J. Woo,$^1$\thanks{Present address: U. Manitoba/TRIUMF, 
Vancouver, B.C. V6T 2A3}  
D.~H. Barkhuff,$^2$\thanks{Present address: 
Massachusetts Institute of Technology, Cambridge, Massachusetts 02139} 
W. Bertozzi,$^3$ 
D. Dale, $^3$\thanks{Present address: University of Kentucky, 
Lexington, Kentucky 40506} G. Dodson,$^3$ K.~A. Dow,$^3$
M.~B. Epstein,$^4$ M. Farkhondeh,$^3$ J.~M. Finn,$^1$
S. Gilad,$^3$ M.~K. Jones,$^{1}$ 
K. Joo,$^3$\thanks{Present address: Thomas Jefferson National Accelerator 
Facility, Newport News, Virginia 23606} 
J.~J. Kelly,$^5$ S. Kowalski$^3$, 
R.~W. Lourie,$^2$\thanks{Present address: State University of New York at 
Stony Brook, Stony Brook, New York 11794} 
R. Madey,$^6$ 
D.~J. Margaziotis,$^4$, P. Markowitz,$^5$\thanks{Present address: Florida 
International University, Miami, Florida 33199}  
J.~I. McIntyre,$^1$\thanks{Present address: Rutgers University, Piscataway, 
New Jersey 08855} 
C. Mertz,$^7$ 
B.~D. Milbrath,$^2$\thanks{Present address: Eastern Kentucky 
University, Richmond, Kentucky 40475}
J. Mitchell,$^8$ C.~F. Perdrisat,$^1$ V. Punjabi,$^{9}$ 
P.~M. Rutt,$^{10}$ 
A.~J. Sarty,$^3$\thanks{Present address: Florida 
State University, Tallahassee, Florida, 32306} 
D. Tieger,$^3$ C. Tschalaer,$^3$ W. Turchinetz,$^3$ P.~E. Ulmer,$^{11}$ 
C. Vellidis,$^{12}$ S.~P. Van Verst,$^{2,3}$\thanks{Present address: Washington
Dept. of Health, Division of Radiation Protection, Olympia, Washington 98504}
G.~A. Warren,$^3$\thanks{Present address: Universit\"{a}t Basel, CH-4056 
Basel, Switzerland} 
L. Weinstein,$^{11}$}

\address{$^1$College of William and Mary, Williamsburg, Virginia 23185\\
         $^2$University of Virginia, Charlottesville, Virginia 22901\\
         $^3$Massachusetts Institute of Technology and Bates Linear Accelerator
Center, Cambridge, Massachusetts 02139\\
         $^4$California State University, Los Angeles, California 90032\\
         $^5$University of Maryland, College Park, Maryland 20742\\
         $^{6}$Kent State University, Kent, Ohio 44242\\
	 $^{7}$Arizona State University, Tempe, Arizona 85287\\
	 $^{8}$Thomas Jefferson National Accelerator Facility, Newport 
News, Virginia 23606\\
	 $^{9}$Norfolk State University, Norfolk, Virginia 23504\\
         $^{10}$Rutgers University, Piscataway, New Jersey 08855\\
         $^{11}$Old Dominion University, Norfolk, Virginia 23529\\
         $^{12}$University of Athens, Greece}

\date{\today}

\maketitle

\begin{abstract}
The first measurements of the induced proton polarization, $P_n$,
for the $^{12}$C$\left(e,e^{\prime}\vec p \right)$ reaction are
reported. The experiment was performed at quasifree kinematics for
energy and momentum transfer $(\omega,q) \approx$ (294 MeV, 756 MeV/c) and
sampled a recoil momentum range of 0-250 MeV/c.
The induced polarization arises from final-state interactions and for these
kinematics is dominated by the real part of the spin-orbit optical potential.
The distorted-wave impulse approximation provides good agreement with
data for the $1p_{3/2}$ shell.
The data for the continuum suggest that both the $1s_{1/2}$ shell and underlying
$\ell > 1$ configurations contribute.
\end{abstract}

\pacs{PAC Numbers:  24.10.Cn,24.10.Eq,24.10.Ht, 24.70.+s,25.30.Dh, 25.30.Fj }
\begin{multicols}{2}[]
\narrowtext

Single-nucleon knockout by electron scattering is sensitive to both 
the nuclear spectral function and to the properties of the electromagnetic
current in the nuclear medium;
recent reviews of this subject may be found in Refs.\
\cite{Kelly96,Boffi93,Boffi96,Dieperink90,FM}.
Single-hole momentum distributions for discrete states of the residual nucleus 
are usually extracted from spin-averaged
differential cross section data. Additional insight into the reaction 
mechanism can be obtained by
separation of the unpolarized response functions.
Even more discriminating tests of the reaction mechanism are provided by
measurements of the polarization of the ejectile.
In this Letter we report the first measurements of recoil polarization for
protons ejected from a nucleus of $A>2$ via electron scattering, 
specifically the $^{12}$C$\left(e,e^{\prime} \vec{p}\right)$ reaction.

Nucleon knockout reactions of the type $A(\vec{e},e^\prime \vec{N})B$
initiated by a longitudinally polarized electron beam and for which the 
ejectile polarization is detected may be described by a differential 
cross section of the form \cite{Giusti89,Picklesimer89}
\begin{equation}
\label{eq:recoil-polarization}
 \frac{d\sigma _{hs}}{d\varepsilon _{f}d\Omega _{e}d\Omega _{N}} =
   \sigma _{0} \frac{1}{2} \left[1 + \bbox{P}\cdot \bbox{\sigma} 
   +h (A + \bbox{P}^\prime \cdot \bbox{\sigma})\right] ,
\end{equation}
where $\varepsilon_f$ is the scattered electron energy, 
$\sigma _{0}$ is the unpolarized cross section, 
$h$ is the electron helicity, 
$s$ denotes the nucleon spin projection upon $\bbox{\sigma}$, 
$\bbox{P}$ is the induced polarization, 
$A$ is the electron analyzing power, and 
$\bbox{P}^{\prime}$ is the polarization transfer.
Thus, the net polarization of the ejectile nucleon $\bbox{\Pi}$ has two 
contributions of the form
\begin{equation}
   \bbox{\Pi} = \bbox{P} + h \bbox{P}^{\prime},
\end{equation}
where $\mid h \mid \leq 1$ is the longitudinal beam 
polarization.
Each of the observables may be expressed in terms of kinematical factors
$V_{\alpha\beta}$ and response functions $R_{\alpha\beta}^{s}$ \cite{Kelly96}.  
The response functions are all bilinear combinations of matrix elements of  
the nuclear electromagnetic current operator.

For coplanar kinematics in which the ejectile momentum lies within the 
electron scattering plane, the induced polarization must be normal to 
the scattering plane while the polarization transfer lies within the 
scattering plane.
Hence, the net ejectile polarization for an unpolarized beam and coplanar
kinematics is normal to the scattering plane.
The recoil polarization is usually calculated with respect to a  
helicity basis in the barycentric frame defined by the basis vectors

\begin{tabular}{lcr} 
$\bbox{\hat{l}} = \frac{\bbox{p}^\prime}{| \bbox{p}^\prime| }$, &
$\bbox{\hat{n}} = \frac{\bbox{q} \otimes \bbox{\hat{l}}}{ |\bbox{q} 
\otimes \bbox{\hat{l}}|}$, &
$\bbox{\hat{t}} = \bbox{\hat{n}} \otimes \bbox{\hat{l}}$  .
\end{tabular}

\noindent
For this experiment performed with ${\bf p}^\prime$ (the 
ejectile three-momentum) on the large-angle
side of ${\bf q}$ (the three-momentum transfer), characterized by      
$\phi_{pq} = 180^{\rm o}$, the $\bbox{\hat{n}}$ direction is vertically 
downwards in the laboratory. 

It can be shown that $P_n$ for the one-photon exchange approximation 
vanishes in the absence of final-state interactions (FSI) between the 
ejectile and the residual system.
Within the distorted-wave impulse approximation (DWIA)
these final-state interactions are usually described by an optical-model 
potential of the form
\begin{eqnarray}
U({\bf r}) &=& U^C(r) + U^{LS}(r) \bbox{\sigma} \cdot \bbox{L} \\
\nonumber
U^C(r) &=& V^C(r) + i W^C(r) \\
\nonumber
U^{LS}(r) &=& V^{LS}(r) + i W^{LS}(r)
\end{eqnarray} 
where $U^C$ and $U^{LS}$ are complex central and spin-orbit potentials,
respectively.
Although the optical potential for elastic scattering from the ground state
can be fit to nucleon-nucleus scattering data, no such
data exists for the
excited states of the residual system that are reached by knockout.
Furthermore, electromagnetic knockout reactions probe the spatial 
distributions of these potentials  
differently than do elastic scattering experiments.
The dynamics of FSI in the continuum may also be
more complicated, requiring explicit channel coupling.
Therefore, $P_n$ provides an important independent test of the optical
model, especially for final states in the continuum. 

The experiment was performed at the MIT-Bates Linear Accelerator Center
using an unpolarized electron beam with an energy of 579 MeV, an average 
current of 25 $\rm \mu$A, and a 1\% duty cycle. 
The carbon target had a thickness of 254 mg/cm$^{2}$.  
Scattered electrons and recoil protons were detected in coincidence 
using the MEPS and OHIPS spectrometers, respectively.  
Both spectrometers consist of two quadrupoles followed by a $90^{\rm o}$ 
vertically-bending dipole (QQD) and are instrumented with vertical drift 
chambers for track reconstruction and scintillation detectors for timing.  
In addition, MEPS contains an Aerogel \v{C}erenkov detector for pion rejection. 

The proton  polarization was measured in the newly commissioned Focal Plane  
Polarimeter (FPP) consisting of a carbon analyzer bracketed by 
two pairs of multi-wire proportional chambers.  
A fast hardware trigger system was used to reject small-angle Coulomb  
scattering events which have small analyzing power \cite{Lourie91b}.
The analyzing power for $120 \leq T_p \leq 200$ MeV was 
measured at 
the Indiana University Cyclotron Facility using proton beams of known 
polarization with this FPP \cite{Lourie93b}. These data were combined with  
the world's 
$p-^{12}C$ analyzing power data for $155 \leq T_p \leq 300$ MeV and 
parameterized in the form introduced by Aprile-Giboni {\it et al.} 
\cite{Aprile-Giboni83}. For this experiment, a 9-cm thick carbon 
analyzer was used which provided an average analyzing power of 0.53.  The  
uncertainty in the measured proton polarization due to the 
analyzing power was 1.6\%.  Details concerning the  spectrometers and the  
FPP can be found elsewhere \cite{McIntyre96,Woo96}.  

The electron spectrometer was set at a scattering angle of 120.3$^{\rm o}$ 
and a central momentum of 280 MeV/c.
The proton spectrometer was set at a central momentum of 756 MeV/c;    
three angle settings (22.03$^{\rm o}$, 26.62$^{\rm o}$, and 31.00$^{\rm o}$) 
were used to cover the missing momentum range $0 \leq p_m \leq 250$ MeV/c.
The ejectile kinetic energy for the ground state of $^{11}$B was 
approximately constant at a central value of 274 MeV and 
$Q^2 = q^2 - \omega^2 = 0.5$ (GeV/c)$^2$.
The data from the three angle settings were combined and binned into 
recoil momentum bins of 50 MeV/c ranging from 0 to 250 MeV/c. 
They were further separated into four missing energy ($E_m$) bins:
A bin from 16.0 to 20.4 MeV where the data were dominated by $1p_{3/2}$ shell  
knockout,
two bins from 28.0 MeV to 39.0 MeV and from 39.0 MeV to 50.0 MeV where the  
reaction is a mixture of $1s_{1/2}$ shell 
knockout and continuum effects, and a bin from 50.0 MeV to 
75.0 MeV where the reaction is primarily due to the continuum.  The measured    
polarizations were corrected for accidental coincidences.  
The signal to noise ratio ranged from 17:1 for $1p_{3/2}$ shell knockout in the 
100-150 MeV/c recoil momentum bin to $\approx$ 1:1 for the 50 $\leq E_m  
\leq$ 75 MeV bins. 

The polarization at the target is related to the polarization at the
focal plane by $P_n^{tgt} = S_{nx} \cdot P^{fp}_x$ where $S_{nx}$ is
a spin-transport coefficient that includes transformations between 
coordinate systems, precession in the magnetic fields, and the effects
of finite acceptance.
For our application, $S_{nx} \approx (\cos \chi_0)^{-1}$, where
$\chi_0 = 207.3^\circ$ is the mean spin-precession angle.
Small corrections for finite acceptance were made by modifying
the Monte Carlo program MCEEP \cite{Ulmer91} to use the spin-transport 
matrices produced by COSY \cite{Berz90}.
The net effect upon $S_{nx}$ varies slowly with $(p_m,E_m)$ and
was found to be in the range $\pm$ 0.03 $\pm$ 0.03,
where the uncertainty includes an estimate of the model dependence
of the Monte Carlo simulation.  
The extracted transvere polarization ($P_t$) averaged over all bins, was 
$P_t=0.008\pm 0.018$.  Also, bin by bin, $P_t$ was consistent with zero, as 
expected for an unpolarized beam.  Instrumental false asymmetries 
for the $P_n$ measurements were shown to be less than $\pm$ 0.005 
from the elastic hydrogen FPP measurement \cite{milmci97}.  
Because the polarization of elastically scattered protons from an unpolarized 
electron beam is constrained to be zero in the one-photon exchange 
limit, any measured polarization provides a means of normalizing the FPP.

\begin{figure}[ht]
\centerline{\strut\psfig{file=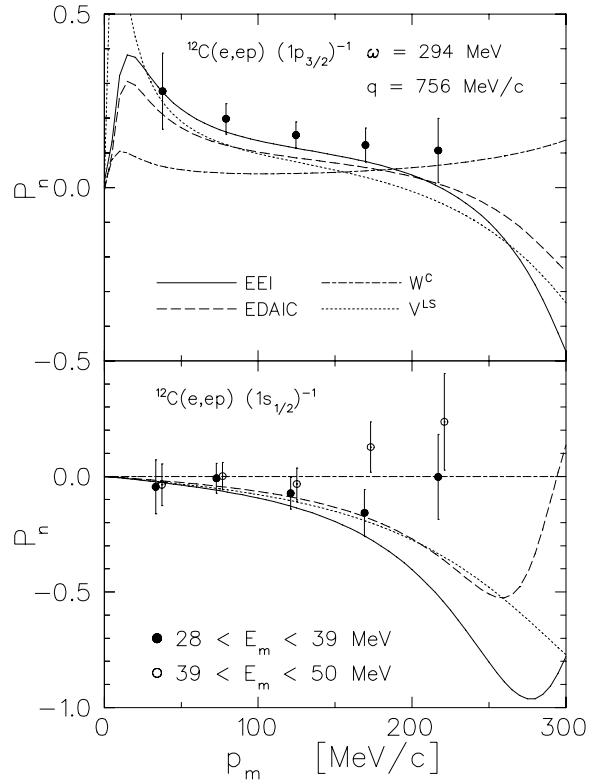,height=4.05in} }
\caption{Polarization for the $^{12}$C$(e,e^\prime \vec{p})$ reaction. 
Data for $E_m \leq 24$ MeV are compared with DWIA calculations for
$1p_{3/2}$ knockout (top).
Data for $28 \leq E_m \leq 39$ MeV and $39 \leq E_m \leq 50$ MeV (bottom) 
are compared with  
calculations for $1s_{1/2}$ knockout, although the relative importance 
of underlying $\ell  > 0$ configurations increases with $p_m$ and 
$E_m$.  Note that a symmetric $\pm$2 MeV/c shift in recoil momentum  
has been put in to separate the data from the two bins.  The solid 
curves show DWIA calculations using an optical potential based upon a 
density-dependent empirical effective interaction (EEI). The long dashed  
curves use the EDAIC potential, whereas other curves show the effect of  
individual components of the optical potential using the EDAIC potential.}
\label{c97fig1}
\end{figure}

\begin{figure}[htb]
\centerline{\strut\psfig{file=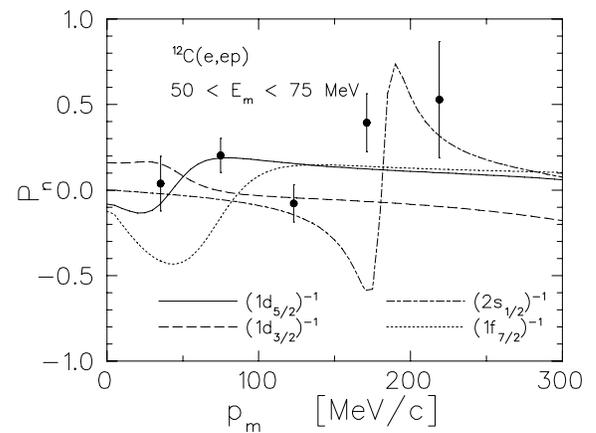,height=2.2in} }
\caption{Polarization for the $^{12}$C$(e,e^\prime \vec{p})$ reaction. 
Data for $50 \leq E_m \leq 75$ MeV are compared with calculations using the  
EDAIC potential for single-nucleon knockout from various orbitals.}
\label{c97fig2}
\end{figure}

Measured polarizations for several bins of missing energy are compared
in Fig.\ \ref{c97fig1} with DWIA calculations using the effective momentum
approximation; 
details of the DWIA formalism may be found in Refs.\ \cite{Kelly96,Kelly97a}.
We used momentum distributions fitted to $^{12}$C$(e,e^\prime p)$ data
by van der Steenhoven {\it et al.}\ \cite{vdSteenhoven88a,vdSteenhoven88b}
and the energy-dependent $^{12}$C optical potential of 
Cooper {\it et al.}\ \cite{Cooper93} with their best fit for carbon 
(EDAIC). 
The Dirac scalar and vector potentials were transformed to equivalent 
Schr\"odinger form and the Darwin nonlocality factor was included.
Fig.\ \ref{c97fig1} shows that DWIA calculations agree reasonably well with the
$P_n$ data for the $1p_{3/2}$ shell with a systematic underestimate of 
about ten percent.
The comparison between DWIA calculations and data for the $1s_{1/2}$ is
complicated by the presence of an underlying continuum that may 
contain significant $\ell > 0$ contributions.
The induced polarization for $28 \leq E_m \leq 39$ MeV is consistent with
DWIA calculations for the $1s_{1/2}$ shell, whereas for $E_m > 50$ 
MeV (Fig.\ \ref{c97fig2}) we find a positive $P_n$.
This result suggests that the polarization of the continuum beneath the
$1s_{1/2}$ shell, composed primarily of configurations with $\ell > 1$, is  
positive and tends to dilute the negative polarization expected for the 
$1s_{1/2}$ shell.   
Thus, the intermediate bin,  $39 \leq E_m \leq 50$ MeV, retains little net
polarization where these opposing contributions tend to
cancel. Note that this effect increases with increasing missing momentum.

The sensitivity to the choice of optical potential is illustrated
in Fig.\ \ref{c97fig1} by comparing calculations based upon the 
EDAIC and EEI optical models.  
The EEI model folds a density-dependent empirical effective 
interaction (EEI) with the nuclear density.
The empirical effective interaction is fitted to proton-nucleus
elastic and inelastic scattering data for several states of several
targets simultaneously using procedures developed in Ref. 
\cite{Kelly89}.
However, because the nearest available energies for the EEI are
200 and 318 MeV \cite{Kelly94}, 
a linear interpolation with respect to ejectile energy was performed.
We find that the EEI model yields somewhat stronger $P_n$
and better agreement with the data for the $1p_{3/2}$ shell. 

It is also instructive to examine the contributions of various components  
of the optical potential separately.
These are illustrated in Fig.\ \ref{c97fig1}
by calculations using the EDAIC potential in which all other parts 
of the optical potential were turned off.
Of course, these separated polarizations do not simply add when the full  
potential is used.
There are two dominant sources of induced polarization: the imaginary
central ($W^C$) and real spin-orbit ($V^{LS}$) potentials.

The most familiar source of induced polarization is produced by $W^C$ and 
arises from the correlation between absorption and initial spin that 
is commonly known as the Newns polarization \cite{Newns53,Newns58} or 
the Maris effect \cite{Jacob76}.
However, spin-orbit distortion is the largest source of induced polarization 
for the present reaction.  Although the effect of spin-orbit 
distortion upon ejectile polarization has been studied for $(d,\vec{p})$ 
reactions at low energies \cite{Butler61,Robson61,Johnson62}, there 
exists little data for the induced polarization in nucleon 
knockout at intermediate energies.  
The nature of the spin-orbit effect can be understood using a semi-classical 
argument \cite{Butler61} based upon the spin-orbit force
\begin{eqnarray}
\bbox{F}^{LS} &=& -\nabla \left( V^{LS}(r) \bbox{\sigma}\cdot \bbox{L} \right) 
\\ \nonumber
&=& -\bbox{\hat{r}} \frac{\partial V^{LS}(r)}{\partial r} 
\bbox{\sigma}\cdot \bbox{L}
+ V^{LS}(r) \bbox{\sigma} \otimes \bbox{p}^\prime \; .
\end{eqnarray}
The first term is a central spin-orbit force which produces spin-correlated
changes in the magnitude of the ejectile momentum and is most important
for parallel kinematics; however, its effect is generally quite small
because it averages over a bipolar function.
The second term is most effective in quasiperpendicular kinematics
where spin-up (spin-down) protons are deflected toward the right (left),
which for $\phi_{pq} = 180^{\rm o}$ increases (decreases) the missing momentum.
The polarization induced by this effect is greatest where the slope
of the initial momentum distribution is largest.  
When $\ell > 0$ a shift of the rising slope of the momentum distribution
toward larger angles for spin-up yields $P_n > 0$, 
whereas the falling slope of an $\ell = 0$ momentum distribution yields
$P_n < 0$ for small $p_m$.
Therefore, this argument explains the sign of $P_n$ for both the
$1p_{3/2}$ and $1s_{1/2}$ states at small $p_m$.  
Furthermore, this argument suggests that zero crossings in the $V^{LS}$  
contribution to $P_n$ should occur near extrema of the momentum distribution, 
but their precise locations depend upon more complicated geometrical 
and refractive effects.

In Fig.\ \ref{c97fig2} we compare the induced polarization for the deep 
continuum, $50 \leq E_m \leq 75$ MeV with DWIA calculations for 
single-nucleon knockout from several orbitals that might be populated
by $2p2h$ ground-state correlations.
Although the overlap functions are not necessarily those of the mean
field, we used Woods-Saxon potentials with depths chosen to reproduce
central missing energy $E_m = 62$ MeV.
For $p_m > 100$ MeV/c we find that knockout from the
$1d_{5/2}$ or $1f_{7/2}$ orbital would produce a positive polarization. 
In addition, the extra node in the $2s_{1/2}$ wave function leads to a
rapid sign change in its contribution to the induced polarization in 
the vicinity of $p_m \sim 180$ MeV/c.
Although this feature will probably be smeared in a more realistic
continuum calculation, a small admixture of this configuration could
have an important effect upon the induced polarization for the continuum
at large $p_m$ where the $1s_{1/2}$ contribution is decreasing rapidly.
Therefore, although more detailed calculations are needed to properly evaluate 
the effect of multinucleon mechanisms both in final-state 
interactions and in the absorption of the virtual photon, it appears 
that single-nucleon knockout from orbitals above the Fermi level that 
would be unpopulated in the absence of two body correlations could 
account for the positive polarization we observe in the deep continuum.

Summarizing, we have performed the first measurements of induced 
polarization for the $^{12}$C$(e,e^\prime \vec{p})$ reaction in 
quasiperpendicular kinematics for $p_m \leq 250$ MeV/c.
The induced polarization is primarily sensitive to final-state interactions
and we have illustrated the roles of each component of the optical potential.
For the present kinematics, the real part of the spin-orbit potential
is the dominant source of $P_n$.
The data for $1p_{3/2}$ shell are in reasonable agreement with standard DWIA
calculations based upon phenomenological optical potentials fit to
elastic scattering data for the ground state. Slightly better agreement with  
the $1p_{3/2}$ shell data is obtained using a density-dependent empirical
effective interaction fitted to proton-nucleus elastic and inelastic
scattering data.  The data for the $1s_{1/2}$ region are also consistent with 
DWIA calculations provided that allowance is made for the opposite 
polarization arising from more complicated contributions to the continuum.
Improved statistical precision should allow the
multipole structure of the continuum and variations of the final-state
interactions for highly excited residual systems to be probed.
Future experiments with polarized electron beams will measure polarization
transfer observables that are expected to be sensitive to two-body currents 
and/or modification of the one-body electromagnetic current,
but relatively insensitive to final-state interactions. Such data,
combined with precise $P_n$ measurements, will result in
considerably more stringent tests of the dynamical ingredients of the
$(\vec{e},e^{\prime} \vec{N})$ process.

\acknowledgements

The authors gratefully acknowledge the work of the staff at MIT-Bates.  
We also thank N.S. Chant for discussions of the history of induced 
polarization in knockout and transfer reactions.  This work was 
supported in part by the U.S. Department of Energy under 
Grant No. DE-FG05-89ER40525 and DE-FG05-90ER40570 and by the National Science  
Foundation under Grants Nos. PHY-89-193959, PHY-91-12816, 
PHY-93-11119, PHY-94-05315, PHY-94-09265, and PHY-94-11620.

\end{multicols}


\begin{thebibliography}{10}

\bibitem{Kelly96}
J.~J. Kelly, Adv.\ Nucl.\ Phys.\ {\bf {\bf 23}},  75  (1996).

\bibitem{Boffi93}
S. Boffi, C. Giusti, and F.~D. Pacati, Phys. Rep. {\bf {\bf 226}},  1  (1993).

\bibitem{Boffi96}
S. Boffi, C. Giusti, F.~D. Pacati, and M. Radici, {\em {Electromagnetic
  Response of Atomic Nuclei}} (Oxford University Press, Oxford, 1996).

\bibitem{Dieperink90}
A.~E.~L. Dieperink and P.~K.~A. de~{Witt Huberts}, {Ann. Rev. Nucl. Part. Sci.}
  {\bf {\bf 40}},  239  (1990).

\bibitem{FM}
S. Frullani and J. Mougey,  in {\em Advances in Nuclear Physics}, edited by
  J.~W. Negele and E. Vogt (Plenum Press, New York, 1984), Vol.~14.

\bibitem{Giusti89}
C. Giusti and F.~D. Pacati, Nucl. Phys. {\bf {\bf A504}},  685  (1989).

\bibitem{Picklesimer89}
A. Picklesimer and J.~W. van Orden, Phys. Rev. {\bf C} {\bf {\bf 40}},  290
  (1989).

\bibitem{Lourie91b}
R.~W. Lourie, S.~P. van Verst, J.~H. Mitchell, D.~H. Barkhuff, and T.~P. Welch,
  Nucl. Instru. Meth. {\bf {\bf A306}},  83  (1991).

\bibitem{Lourie93b}
R.~W. Lourie {\it et~al.}, Technical report, IUCF Annual and Technical Report
  (unpublished).

\bibitem{Aprile-Giboni83}
E. Aprile-Giboni {\it et~al.}, Nucl. Instru. Meth. {\bf {\bf 215}},  147
  (1983).

\bibitem{McIntyre96}
J.~I. McIntyre, Ph.D. thesis, The College of William and Mary, 1996.

\bibitem{Woo96}
R.~J. Woo, Ph.D. thesis, The College of William and Mary, 1996.

\bibitem{Ulmer91}
P.~E. Ulmer, Technical Report No.~{CEBAF-TN-91-101}, {CEBAF} (unpublished).

\bibitem{Berz90}
M. Berz, Nucl. Instru. Meth. {\bf {\bf A298}},  364  (1990).

\bibitem{milmci97}
B.~D. Milbrath, J.~I. McIntyre {\it et~al.}, to be published.

\bibitem{Kelly97a}
J.~J. Kelly, Phys. Rev. {\bf C} {\bf 56}, 2672 (1997).

\bibitem{vdSteenhoven88a}
G. van~der Steenhoven, H.~P. Blok, E. Jans, M. de~Jong, L. Lapik\'{a}s,
  E.~N.~M. Quint, and P.~K.~A. de~Witt~Huberts, Nucl. Phys. {\bf {\bf A480}},
  547  (1988).

\bibitem{vdSteenhoven88b}
G. van~der Steenhoven, H.~P. Blok, E. Jans, L. Lapik\'{a}s, E.~N.~M. Quint, and
  P.~K.~A. de~Witt~Huberts, Nucl. Phys. {\bf {\bf A484}},  445  (1988).

\bibitem{Cooper93}
E.~D. Cooper, S. Hama, B.~C. Clark, and R.~L. Mercer, Phys. Rev. {\bf C} {\bf
  {\bf 47}},  297  (1993).

\bibitem{Kelly89}
J.~J. Kelly, Phys. Rev. {\bf C} {\bf 39}, 2120 (1989)

\bibitem{Kelly94}
J.~J. Kelly and S. J. Wallace, Phys. Rev. {\bf C} {\bf 49}, 1315 (1994)

\bibitem{Newns53}
H.~C. Newns, Proc. Phys. Soc. London {\bf {\bf A66}},  477  (1953).

\bibitem{Newns58}
H.~C. Newns and M.~Y. Refai, Proc. Phys. Soc. London {\bf {\bf A71}},  627
  (1958).

\bibitem{Jacob76}
G.~Jacob, T.~A.~J. Maris, C. Schneider, and M.~R. Teodoro, Nucl. Phys. {\bf
  {\bf A257}},  517  (1976).

\bibitem{Butler61}
S.~T. Butler, in {\em Proc. of the Rutherford Jubilee International 
Conference}, edited by J.~B. Birks, (Heywood and Company, London, 1961), p. 492,
L.~J.~B. Goldfarb, {\it loc cit.}, p. 479.

\bibitem{Robson61}
D. Robson, Nucl. Phys. {\bf 22}, 34 (1961).

\bibitem{Johnson62}
R.~C. Johnson, Nucl. Phys. {\bf 35}, 654 (1962).

\end{thebibliography}
\end{document}